\pdfoutput=1

\documentclass[english,aps,prd,superscriptaddress,preprintnumbers,floatfix,nofootinbib,12pt]{revtex4-1}

\usepackage{amsmath,amssymb,amsfonts}
\usepackage{amsmath}
\usepackage{graphicx}
\usepackage{comment}
\usepackage{mathtools}
\usepackage{slashed}
\usepackage{physics}
\usepackage{mathrsfs}
\usepackage{enumerate}

\usepackage{bm}
\usepackage[utf8]{inputenc}

\usepackage{listings}

\usepackage[T1]{fontenc}
\usepackage{lmodern}

\usepackage{hyperref}

\usepackage{dsfont}

\hypersetup{colorlinks,linkcolor={blue},citecolor={blue},urlcolor={blue}} 
\usepackage{braket}
\usepackage[english]{babel}
\usepackage{bm}
\usepackage[utf8]{inputenc}

\usepackage{listings}

\usepackage[T1]{fontenc}
\usepackage{lmodern}
\usepackage{caption}

\usepackage{floatrow}

\usepackage[normalem]{ulem}
\usepackage{color}
\usepackage{array}
\def\beq{\begin{equation}}
\def\eeq{\end{equation}}
\def\be{\begin{equation}}
\def\ee{\end{equation}}
\def\bea{\begin{eqnarray}}
\def\eea{\end{eqnarray}}

\begin{document}

\title{Neutrino Lorentz Invariance Violation from Cosmic Fields}   

\author{Rub\'en Cordero}\email{rcorderoe@ipn.mx} \affiliation{Departamento de F\'{\i}sica, Escuela Superior de
  F\'{\i}sica y Matem\'aticas del Instituto Polit\'ecnico Nacional, Unidad Adolfo L\'opez Mateos, Edificio 9, 07738 Ciudad de M\'exico, Mexico}
\author{Luis A. Delgadillo} \email{ldelgadillof2100@alumno.ipn.mx}
\affiliation{Departamento de F\'{\i}sica, Escuela Superior de
  F\'{\i}sica y Matem\'aticas del Instituto Polit\'ecnico Nacional, Unidad Adolfo L\'opez Mateos, Edificio 9, 07738 Ciudad de M\'exico, Mexico}

\begin{abstract}
\noindent
From a cosmological perspective, scalar fields are well-motivated dark matter and dark energy candidates. Several possibilities of neutrino couplings with a time-varying cosmic field have been investigated in the literature. In this work, we present a framework in which violations of Lorentz invariance (LIV) and $CPT$ symmetry in the neutrino sector could arise from an interaction among neutrinos with a time-varying scalar field. Furthermore, some cosmological and phenomenological aspects and constraints concerning this type of interaction are discussed. Potential violations of Lorentz and $CPT$ symmetries at present and future neutrino oscillation experiments such as IceCube and KM3NeT can probe this scenario. 

\end{abstract}

\maketitle




\section{Introduction}
\label{intro}

Possible Lorentz invariance violations (LIV) and violations of the $CPT$ symmetry in the neutrino sector could arise from the interactions among scalar fields with neutrinos via exotic currents \cite{Gu:2005eq, Ando:2009ts, Klop:2017dim, Capozzi:2018bps}. For comprehensive treatments of Lorentz and $CPT$ violations in neutrinos, we refer to Refs.~\cite{Kostelecky:2003cr, Diaz:2016xpw, Moura:2022dev}. In a model-independent framework, LIV effects are parameterized by the effective Lagrangian~\cite{Barenboim:2022rqu}
\begin{equation}
    \mathcal{L}_{ \text{eff}}^{\text{Fermion}} = \mathcal{L}_{\text{SM}}^{\text{Fermion}} + \mathcal{L}_{\text{LIV}} + \text{h.c.},
\end{equation}

\begin{equation}
    \mathcal{L}_{\text{LIV}}= - \frac{1}{2} \Big\{ p^{\mu}_{\alpha \beta}\Bar{\psi}_{\alpha} \gamma_{\mu} \psi_{\beta} + q^{\mu}_{\alpha \beta}\Bar{\psi}_{\alpha} \gamma_{5}  \gamma_{\mu}  \psi_{\beta} -i r^{\mu \nu}_{\alpha \beta}\Bar{\psi}_{\alpha} \gamma_{\mu}  \partial_{\nu}  \psi_{\beta}- i s^{\mu \nu}_{\alpha \beta}\Bar{\psi}_{\alpha}  \gamma_{5}\gamma_{\mu}  \partial_{\nu} \psi_{\beta} \Big\}~,
\end{equation}

in the case of neutrinos, it is convenient to define
\begin{equation}
    (a_{L})^{\mu}_{\alpha \beta} = (p+q)^{\mu}_{\alpha \beta}, ~~~ (c_{L})^{\mu \nu}_{\alpha \beta} = (r+s)^{\mu \nu }_{\alpha \beta},
\end{equation}
the $(a_{L})^{\mu}_{\alpha \beta}$ coefficients are $CPT$-odd, while the $ (c_{L})^{\mu\nu}_{\alpha \beta}$ coefficients are $CPT$-even.
Besides, if we are interested to study the isotropic LIV contributions at neutrino oscillation experiments, namely the $CPT$-even ($c^{00}_{\alpha \beta}=c_{\alpha \beta}$) and $CPT$-odd ($a^{0}_{\alpha \beta}=a_{\alpha \beta}$) coefficients, i.e. $(\mu = \nu = 0)$, we can consider the following Hamiltonian
\begin{equation}
    H=H_{\text{vacuum}} + H_{\text{matter}} + H_{\text{LIV}},
\end{equation}
where the contribution from the LIV sector is
\begin{equation}
\label{LIVHAM}
 H_{\text{LIV}} =   \left(
\begin{array}{ccc}
 a_{ee} & a_{e \mu} & a_{e \tau} \\
 a_{e \mu}^{*} & a_{\mu \mu} &  a_{ \mu \tau} \\
 a_{e \tau}^{*} & a_{ \mu \tau}^{*} & a_{ \tau \tau} \\
\end{array}
\right) - \frac{4}{3} E \left(
\begin{array}{ccc}
 c_{ee} & c_{e \mu} & c_{e \tau} \\
 c_{e \mu}^{*} & c_{\mu \mu} &  c_{ \mu \tau} \\
 c_{e \tau}^{*} & c_{ \mu \tau}^{*} & c_{ \tau \tau} \\
\end{array}
\right)\,\, ,
\end{equation}
being $E$ the neutrino energy, the LIV coefficients $(a_{L})^{\mu}_{\alpha \beta}$ also violate $CPT$ in the neutrino sector. From a theoretical point of view, violations of the $CPT$ symmetry manifest themselves within the framework of string theory~\cite{Kostelecky:1988zi, Kostelecky:1991ak, Kostelecky:1995qk}. Regarding neutrino oscillation experiments, $CPT$ violations in the neutrino sector were proposed as a solution to the Liquid Scintillator Neutrino Detector (LSND) anomaly~\cite{Murayama:2000hm, Barenboim:2001ac, Barenboim:2004wu}. Furthermore, it was shown in Ref.~\cite{Barenboim:2002hx} that besides neutrino oscillation experiments, $CPT$ violation can have consequences in neutrinoless double-beta decay experiments.

Recently, it has been considered the case where violations of Lorentz invariance and $CPT$ at neutrino oscillation experiments,~\footnote{An early model for $CPT$ violation in the neutrino sector was discussed in~\cite{Barenboim:2002tz}.} arise from a neutrino current coupled with a time-dependent cosmic field $\phi(t)$ (see e.g., Refs.~\cite{Ando:2009ts, Klop:2017dim, Capozzi:2018bps})
\begin{equation}
\label{livdark}
    a^{\mu}_{\alpha \beta} \Bar{\nu}_{\alpha} \gamma_{\mu}(1- \gamma_5) \nu_{\beta} \rightarrow \lambda_{\alpha \beta} \frac{\partial^{\mu} \phi}{\Lambda}  \Bar{\nu}_{\alpha} \gamma_{\mu}(1- \gamma_5) \nu_{\beta},
\end{equation}
being $\lambda_{\alpha \beta}$ some coupling constant, $\Lambda$ the energy scale of the interaction, and $\phi$ a scalar field, which could be identified as a dark matter (DM) or dark energy (DE) candidate. Here, violations of Lorentz invariance emerge via $CPT$-odd LIV coefficients $a_{\alpha \beta}^{\mu} \rightarrow \lambda_{\alpha \beta} \partial^{\mu} \phi \Lambda^{-1}$. 

In the scenario explored by the authors of Refs.~\cite{Gu:2005eq, Ando:2009ts, Klop:2017dim}, the scalar field $\phi$ is considered to be a DE candidate (namely quintessence) 
\begin{equation}
   \dot{\phi}_0 \simeq  H_0  M_{\text{Pl}} (1+w)^{1/2},
\end{equation}
being $H_0 \sim 10^{-33}$ eV the Hubble parameter at present, $M_{\text{Pl}}$ the Planck mass, and the equation of state parameter $w \simeq -0.95$ \cite{Planck:2018vyg}. Hence, $\dot{\phi}_0 \sim H_0M_{\text{Pl}} (1-0.95)^{1/2} \sim 10^{-24}$ GeV$^2$. Furthermore, the isotropic $CPT$-odd LIV coefficients are related to the scalar field $\phi$ as
\begin{equation}
    |a_{\alpha \beta}| \sim |\lambda_{\alpha \beta}| \frac{\dot{\phi}(t)}{\Lambda}. 
\end{equation}
For instance, the expected sensitivity to the isotropic LIV coefficients at DUNE is $|a_{\alpha \beta}| \sim 10^{-23}$ GeV~\cite{Barenboim:2018ctx, Agarwalla:2023wft}. Therefore, for couplings $|\lambda_{\alpha \beta}| \sim \mathcal{O}(1)$, a DUNE-like experiment ($|a_{\alpha \beta}| \sim 10^{-23}$ GeV) could probe a mediator mass scale up to $\Lambda \sim 10^2$ MeV $\ll M_{W} \sim 10^2$ GeV. Moreover, in this scenario, experiments such as IceCube and the KM3NeT proposal could search for directional-dependent LIV effects at neutrino energies $E_\nu \gtrsim 10^5$ GeV~\cite{Klop:2017dim, Telalovic:2023tcb}, the sensitivity reach to the mediator mass scale $\Lambda$ would be $\Lambda \lesssim 10^5$ GeV or $\Lambda \lesssim 10^8$ GeV, depending on the neutrino event exposure (see, e.g., Fig.~5a and Fig.~5b of Ref.~\cite{Klop:2017dim} for a detailed explanation).  

On the other hand, if $\phi$ is a DM candidate, for instance, ultralight dark matter (ULDM) with a parabolic potential $V(\phi) \simeq m_\phi^2 \phi^2$~(see, e.g.~\cite{Magana:2012ph, Suarez:2013iw, Ferreira:2020fam} and references therein)
\begin{equation}
   \dot{\phi}_{\odot} \simeq  m_{\phi} \phi_{\odot} \big(1+|\langle v_\phi \rangle \hat{v}_\phi \cdot \hat{p}_\nu | \big)~,
\end{equation}
by considering a boson mass $m_{\phi}  \sim 10^{-20}$ eV, with local density $\rho_{\phi, \odot} \sim 10^{5} \rho_{\text{DM}}$~\cite{Planck:2018vyg},~\footnote{From recent analyses, the estimation of the local DM density coincides within a range of $\rho_{\text{DM}, \odot} \simeq 0.3-0.6~ \text{GeV/cm}^3$~\cite{deSalas:2019pee,deSalas:2020hbh,Sivertsson:2022riu}.} and neutrino propagation direction $\hat{p}_\nu$, providing the virialized ULDM velocity in the Milky Way is of the order $\langle v_\phi \rangle \sim \mathcal{O}(10^{-3})c$, the ULDM amplitude at present would give $\phi_{\odot} \sim 10^{17}$ eV, and $\dot{\phi}_{\odot} \sim 10^{-3}$ eV$^2\sim 10^{-21}$ GeV$^2$. Therefore, for couplings $|\lambda_{\alpha \beta}| \sim \mathcal{O}(1)$, DUNE ($|a_{\alpha \beta}| \sim 10^{-23}$ GeV) could search for a mediator mass scale up to $\Lambda \sim 10^2$ GeV.

\section{Theoretical framework}
\label{model}
Let us consider the following phenomenological Lagrangian~\footnote{Neutrino oscillations in the presence of an effective interaction of this type, can be employed to test the Earth's dark matter halo~\cite{Gherghetta:2023myo}. A similar interaction was studied in Refs.~\cite{Simpson:2016gph, Burgess:1993xh} in the context of majoron-neutrino coupling. Furthermore, in Ref.~\cite{Kitazawa:2022gzk} the scalar field couples to a fermion current via a non-linear Callan-Coleman-Wess-Zumino (CCWZ)~\cite{Coleman:1969sm, Callan:1969sn} effective Lagrangian.}
\begin{equation}
\label{uldmn}
    \mathcal{L}_{\phi, \nu} = g_{\alpha \beta} \frac{\partial^{\mu} \phi}{M_{Z^{\prime}}}  \Bar{\nu}_{\alpha} \gamma_{\mu}(1- \gamma_5) \nu_{\beta}, 
\end{equation}
which can emerge, for instance, from the interaction among a real scalar field $\phi$ and a new vector boson $Z^{\prime}_\mu$ in a broken approximate global symmetry~\footnote{For instance, a scalar field $\phi$ that possesses an approximate shift symmetry $\phi \simeq \phi+2 \pi$~\cite{Hui:2016ltb} (see Appendix~\ref{reali} for a possible realization).}
\begin{equation}
    \label{uldmz}
    \mathcal{L}_{\phi, Z^{\prime}}  \supset M_{Z^{\prime}} \partial^{\mu} \phi Z^{\prime}_\mu,
\end{equation}
and a neutrino current
\begin{equation}
\label{neutrinoz}
  \mathcal{L}_{\nu, Z^{\prime}} \supset  g_{\alpha \beta} \Big( \Bar{\nu}_{\alpha} \gamma^{\mu} (1-\gamma^5) \nu_{\beta} \Big) Z^{\prime}_{\mu},
\end{equation}
where $g_{\alpha \beta}$ is the $3\times3$ coupling matrix. Hence, by integrating out the \emph{neutrinophilic} $Z^{\prime}_{\mu}$ boson from Eqs.~(\ref{uldmz}) and (\ref{neutrinoz}) we can obtain Eq.~(\ref{uldmn}). There exist some models in which the neutrinos are the only standard model fermions that couple to the $Z^{\prime}_\mu$ boson at tree level (see, e.g.~\cite{Bakhti:2018avv, Kelly:2020pcy, Abdallah:2021npg} and references therein). 

For instance, it might be feasible to search for either a $200-500$ GeV neutrinophilic $Z^{\prime}_{\mu}$ boson with couplings $g_{\alpha \beta} \sim 10^{-2}$ via $Z-Z^{\prime}$ mixing ($\theta_{ZZ^{\prime}} \lesssim 0.1 g_{\alpha \beta}$) at the LHC~\cite{Abdallah:2021npg} or a $10-100$ GeV neutrinophilic vector boson with couplings $g_{\alpha \beta} \lesssim 10^{-3}-10^{-2}$ via Higgs decay ($h\rightarrow \nu \Bar{\nu} Z^{\prime}$)~\cite{Kelly:2020pcy}, corresponding to the effective couplings $g_{\alpha \beta}/M_{Z^{\prime}} \lesssim 10^{-13}$ eV$^{-1}$, which could be relevant for searches of Lorentz-invariance violations at present and future neutrino oscillation experiments via the $CPT$-odd coefficients ($a^{\mu}_{\alpha \beta} \rightarrow g_{\alpha \beta} \partial^{\mu} \phi / M_{Z^{\prime}}$). Notice that the effective couplings $g_{\alpha \beta}/M_{Z^{\prime}} \lesssim 10^{-13}$ eV$^{-1}$ remain with respect to an $\mathcal{O}(\text{TeV})$ $Z^{\prime}_{\mu}$ boson with couplings $g_{\alpha \beta} \lesssim 0.1$. Therefore, a boson mass $M_{Z^{\prime}} \sim \mathcal{O}(10~\text{GeV})-\mathcal{O}(\text{TeV})$ will be of phenomenological interest within this framework.

Besides, a related scenario of neutrino-ULDM interaction was proposed by~\cite{Farzan:2018pnk}, the model preserves the flavor composition of high-energy astrophysical neutrinos at production; it includes a complex ultralight scalar field as a DM candidate and a light $Z^{\prime}_{\mu}$ boson. Recently, the authors of Ref.~\cite{Davoudiasl:2023uiq} investigated the interaction of an $\mathcal{O}$(eV) sterile neutrino coupled with an ultralight scalar field $m_{\phi} \sim 10^{-15}$ eV as DM.

\section{Cosmology}
\label{cosmo}
The cosmological evolution of the interacting scalar field with  neutrinos can be described through energy exchange between the two fluids~\cite{Boehmer:2008av, Chongchitnan:2008ry}
\begin{equation}
    \dot{\rho}_{\nu} +4H\rho_\nu =Q,
\end{equation}
\begin{equation}
    \dot{\rho}_{\phi} +3H(\rho_\phi + P_\phi) =-Q,
\end{equation}
where the explicit form of the energy exchange term $Q$ depends on the model. In some models~\cite{Simpson:2016gph}, the explicit form of this term is
\begin{equation}
    Q= \tilde{g}n_\nu \ddot{\phi} \, ,
\end{equation}
where $\tilde{g}$ is a coupling constant. However, $Q\propto \dot{\phi}$~\cite{Boehmer:2008av, Chongchitnan:2008ry} is a possible scenario as well.

The evolution of the scalar field can be determined from its equation of motion
\begin{equation}
\ddot{\phi}+3H\dot{\phi}+\frac{dV(\phi)}{d\phi}= -\frac{Q}{\dot{\phi}},
    \label{fievolution}
\end{equation}
where $H = \dot{a}/a$ is the Hubble parameter and $a$ is the scale factor of the Universe. The evolution of the Hubble parameter can be obtained from the Friedmann equation
\begin{equation}
    H^2 = \frac{8\pi G}{3}(\rho_r+ \rho_b+ \rho_{\text{DM}} + \rho_{\text{DE}}).
\end{equation}
Here, $\rho_r$, $\rho_b$, $\rho_{\text{DM}}$, and $\rho_{\text{DE}}$ are the energy densities associated with radiation, nonrelativistic baryonic matter, dark matter, and dark energy, respectively. In some cases, we can identify $\rho_\phi$ as $\rho_{\text{DM}}$ and in another case as $\rho_{\text{DE}}$ (see below).

\subsection{Neutrino$-$scalar field back-reaction}

\subsubsection{Scalar field as dark energy}
Besides, the equation of motion for the scalar field is modified by the presence of the neutrino$-$scalar field energy exchange. If we consider the scenario where $Q\propto\ddot{\phi}$~\cite{Simpson:2016gph}, we obtain
\begin{equation}
\label{eqmotint}
    \ddot{\phi}\Big(1+ \frac{g_{\alpha \beta} n_\nu }{M_{Z^{\prime}} \dot{\phi}}  \Big) + 3H \dot{\phi} + \frac{d V(\phi, \dot{\phi})}{d \phi} =0.
\end{equation}
The evolution of the field may tend to be constant near the region when the derivative of the scalar field is null.~\footnote{See Appendix \ref{scalarevo} for a discussion.} This could imply avoiding the oscillations of the field and preventing it from being considered DM. It would be interesting to study a possible mechanism to start an oscillatory behavior by perturbing the field or considering quantum fluctuations.

The scenario when the scalar field could be frozen (when $|\dot{\phi}| \ll g_{\alpha \beta} n_\nu  M_{Z^{\prime}}^{-1}$) at some point and behave as DE was studied in~\cite{Simpson:2016gph}. Furthermore, the phenomenology of the scalar field $\phi$ as DE, LIV and $CPT$ violations in neutrino oscillations was studied in Refs.~\cite{Gu:2005eq, Ando:2009ts, Klop:2017dim}.

\subsubsection{Scalar field as dark matter}
Regarding the scalar field $\phi$, if we consider it to be a DM candidate, namely ultralight dark matter, with potential $V(\phi) \simeq m_\phi^2 \phi^2$, the corresponding field is~\cite{Suarez:2013iw, Ferreira:2020fam}
\begin{equation}
   \phi(t,X) \simeq \frac{\sqrt{2 \rho_{\phi, \odot}}}{m_\phi} \sin\big(m_\phi (t -\langle v_\phi \rangle X) \big)~,
\end{equation}
being $\rho_{\phi, \odot}$ the scalar field density and $\langle v_\phi \rangle$ the virialized ULDM velocity in the Milky Way (MW), which is much smaller than the speed of light $\langle v_\phi \rangle\sim 10^{-3}c$, and $X=(x,~y,~z)$ are the spatial components. As reference values, let us consider a scalar field density $\rho_{\phi} \sim  \rho_{\text{DM}} \sim 10^{-11}$ eV$^4$ (locally $\rho_{\text{DM}, \odot} \sim 10^{5} \rho_{\text{DM}}$~\cite{Planck:2018vyg}), with mass $m_\phi \simeq 10^{-15}$ eV.~\footnote{If the scalar field mass is lighter than about $m_\phi \lesssim 10^{-20}$ eV, the scalar field density $\rho_{\phi}$ is not allowed to be the total component of cosmological DM $\rho_{\text{DM}}$, see e.g. Ref.~\cite{Cordero:2022fwb}.} For instance, several ultralight scalars (down to $m\sim 10^{-33}$ eV) appear in some string theories~\cite{Arvanitaki:2009fg,Cicoli:2021gss}. As an exemplification, within the $M$-theory axiverse~\cite{Acharya:2010zx,Marsh:2015xka} an ultralight ($m \sim 10^{-15}$ eV) axion-like particle arises, with decay constant $F\sim 10^{16}$ GeV. \footnote{Searches of ultralight scalars as dark matter candidates include atomic clocks~\cite{Arvanitaki:2014faa}, resonant-mass detectors~\cite{Arvanitaki:2015iga} and atomic gravitational wave detectors~\cite{Arvanitaki:2016fyj}.}

During the early epochs, an effective back-reaction potential $V(\dot{\phi})$ develops due to the neutrino$-$scalar field interaction in Eq.~(\ref{uldmn}), where $V(\dot{\phi}) \lesssim V(\phi) \sim m_\phi^2 \phi^2$ in such way that the field $\phi \simeq \phi(t)$ behaves as DM. For instance, prior to matter-radiation equality ($T\sim$ eV), the Universe was highly isotropic and neutrinos were relativistic. Thus, the corresponding average of the neutrino current was $\langle \Bar{\nu}_{\alpha} \gamma^{\mu}(1- \gamma^5) \nu_{\beta} \rangle \sim \langle \Bar{\nu}_{\alpha} \gamma^{0}(1- \gamma^5) \nu_{\beta} \rangle \sim n_\nu$, with neutrino number density proportional to the temperature $n_\nu\sim 0.1~T^3$, therefore the effective back-reaction potential was
\begin{equation}
 V(\dot{\phi}) \simeq \frac{g_{\alpha \beta} n_\nu}{M_{Z^{\prime}}} \dot{\phi}.
\end{equation}
For the case of a non-interacting scalar field $\phi$ with mass $m_\phi \sim 10^{-15}$ eV, it will transition to its oscillation phase around a temperature $T_{\text{osc}, \phi} \sim \sqrt{ M_{\text{Pl}} m_{\phi} } \sim$ MeV~\footnote{For a derivation of this relationship, we refer the reader to Sec.~II of Ref.~\cite{Hui:2016ltb}.} ($z_{\text{osc}} \sim 6\times 10^9$), the corresponding scalar field amplitude around that time was roughly $|\phi|  \sim 10^{24} ~\text{eV}$, with potential  
\begin{equation}
 | V(\phi) |  \sim m_\phi^2 | \phi^2  |  \sim 10^{-6} \text{MeV}^4,
\end{equation}
therefore, the contribution from the effective back-reaction potential was 
\begin{equation}
|V(\dot{\phi})| \sim  \frac{g_{\alpha \beta} n_\nu}{M_{Z^{\prime}}}| \dot{\phi}| \sim 10^{-11} \text{MeV}^4 ~(T/\text{MeV})~(g_{\alpha \beta} /0.1)~(|\dot{\phi}|/10^{-3} \text{MeV}^2)~(M_{Z^{\prime}}/ 10^{12} \text{eV}), 
\end{equation}
hence at that epoch, the back-reaction potential was a sub-leading contribution to the scalar field potential $|V(\dot{\phi})| \sim  10^{-5} | V(\phi) |$. 

In contrast, if we consider an energy exchange term, $Q = g_{\alpha \beta} n_\nu  M_{Z^{\prime}}^{-1} m_\nu  \dot{\phi} \simeq m_\nu V(\dot{\phi})$ (with neutrino mass $m_\nu\sim 0.1$ eV), we obtain the following equation of motion\begin{equation}
\label{eqmotint2}
    \ddot{\phi} + 3H \dot{\phi} + \frac{d V(\phi, \dot{\phi})}{d \phi} + \frac{g_{\alpha \beta} n_\nu  m_\nu}{M_{Z^{\prime}} }=0.
\end{equation}
Thus, the energy exchange term ($Q \propto \dot{\phi}$) has to satisfy
\begin{equation}
\label{condi}
    \frac{g_{\alpha \beta} n_\nu  m_\nu}{M_{Z^{\prime}} }\lesssim \Big|\frac{d V(\phi, \dot{\phi})}{d \phi}\Big|,
\end{equation}
in order for $\phi$ to start its rapid oscillations and act as DM. For instance, around matter-radiation equality ($T\sim$ eV), an order of magnitude estimate gives
\begin{equation}
\label{esti}
    \frac{g_{\alpha \beta} n_\nu  m_\nu}{M_{Z^{\prime}} }\sim 10^{-15} \text{eV}^3~(g_{\alpha \beta} /0.1)~(M_{Z^{\prime}}/ 10^{12} \text{eV}) \lesssim m_\phi^2 |\phi| \sim 10^{-15} \text{eV}^3~(m_\phi/10^{-15} \text{eV})~(|\phi|/10^{15} \text{eV}).
\end{equation}
In addition, once neutrinos enter the nonrelativistic regime ($T\lesssim m_\nu$), the spatial terms from the neutrino$-$scalar field interaction in Eq.~(\ref{uldmn}) become relevant, modifying the form of Eqs.~(\ref{eqmotint}) and (\ref{eqmotint2}).

On the other hand, in terms of the background quantities, one could have either $Q=\chi \kappa \rho_\nu \dot{\phi}$ or $Q=\chi H\rho_\nu$ (with $\chi$ a dimensionless constant, see e.g. Ref.~\cite{Boehmer:2008av}) which leads to
\begin{equation}
\label{eqmotint3}
    \ddot{\phi} + 3H \dot{\phi} + \frac{d V(\phi, \dot{\phi})}{d \phi} + \chi \Big(\kappa \rho_\nu~\text{or}~\frac{H \rho_\nu}{\dot{\phi}} \Big)=0,~~\kappa^2=\frac{8 \pi G}{3}\sim \frac{1}{M_{\text{Pl}}^2},
\end{equation}
in the case of energy exchange $Q\propto \rho_\nu M_{\text{Pl}}^{-1}$, the neutrino-scalar field interaction is suppressed by the Planck mass. However, the last term with energy exchange $Q=\chi H\rho_\nu$ might be problematic if we consider $\phi$ as DM candidate. 

Hence, if the energy exchange term, $Q \propto \ddot{\phi}$, the scalar field could freeze at some point and behave as DE~\cite{Simpson:2016gph}, while if the energy exchange term, $Q \propto \dot{\phi}$, the scalar field could avoid freezing and act as DM. Besides, in the case where the energy exchange among the neutrinos and the scalar field is $Q\propto \dot{\phi}$, the beginning of the rapid scalar field oscillations can be delayed (see Eqs.~\ref{condi} and \ref{esti}). Therefore, as long as the fast oscillations occurred after BBN but before the CMB formation, $\phi$ can account for all DM~\cite{Krnjaic:2017zlz, Cordero:2022fwb}.

Henceforth, we will discuss some consequences for neutrino oscillation experiments in the case where the scalar field $\phi$ can be considered a DM candidate.

\section{Implications in neutrino oscillation experiments}
\label{implications}
The phenomenological Lagrangian in Eq.~(\ref{uldmn}) can induce violations of Lorentz invariance and $CPT$ at neutrino oscillation experiments via the $CPT$-odd LIV coefficients $a^{\mu}_{\alpha \beta} \rightarrow g_{\alpha \beta} \partial^{\mu} \phi / M_{Z^{\prime}}$ ($g_{\alpha \beta}/M_{Z^{\prime}} \lesssim 10^{-13}$ eV$^{-1}$). Moreover, considering $\phi$ as ULDM with mass $m_\phi \sim 10^{-15}$ eV; thus, the local value of the temporal component of the scalar field amplitude (at present) in the MW is $|\phi_{\odot}| \sim 10^{12}$ eV, therefore $|\dot{\phi}_{\odot}| \simeq m_\phi |\phi_{\odot}| \simeq \sqrt{2\rho_{\phi,\odot}} \sim 10^{-21}$ GeV$^2$, where the isotropic LIV coefficients $a_{\alpha \beta}(t)$ are
\begin{equation}
    a_{\alpha \beta} (t) = a_{\alpha \beta} \cos(m_\phi t) \simeq \frac{g_{\alpha \beta} |\dot{\phi}_{ \odot}| }{M_{Z^{\prime}} } \cos(m_\phi t),
\end{equation}
with $a_{\alpha \beta} \lesssim 10^{-25}$ GeV at our benchmark values. In a similar fashion as the scenario explored in Ref.~\cite{Klop:2017dim}, directional-dependent effects ($a_{\alpha \beta}^{X}$) will be sub-leading due to the small anisotropic components of the scalar field amplitude, which are proportional to the virialized ULDM velocity $\langle v_\phi \rangle \sim 10^{-3}$,   
\begin{equation}
    |a_{\alpha \beta}^{X}|\simeq \frac{g_{\alpha \beta} }{M_{Z^{\prime}} }|\nabla \phi_{\odot} \cdot \hat{p}_\nu| \simeq \frac{g_{\alpha \beta} }{M_{Z^{\prime}} } m_\phi |\phi_{\odot} \langle {v}_\phi\rangle \hat{v}_\phi \cdot \hat{p}_\nu| \sim \frac{g_{\alpha \beta} }{M_{Z^{\prime}} } \sqrt{2\rho_{\phi,\odot}} \langle {v}_\phi\rangle \lesssim 10^{-28}~\text{GeV},
\end{equation}
these coefficients have been studied in the context of active-sterile neutrino oscillations and supernovae neutrino emission~\cite{Lambiase:2023hpq}. 

Imprints of the aforementioned Lorentz$/CPT$ effects can be traced in neutrino oscillation experiments. For instance, the effective neutrino Hamiltonian $\tilde{H}(t)$ in the presence of the $CPT$-odd LIV coefficients is
\begin{equation}
    \tilde{H}(t) = H_0 +V_{\text{MSW}} +a_{\alpha \beta}(t)+a_{\alpha \beta}^{X}(t),
\end{equation}
where $H_0$ is the neutrino Hamiltonian in vacuum, $V_{\text{MSW}}$ is the MSW potential, and the last terms correspond to the $CPT$-odd LIV Hamiltonian $a_{\alpha \beta}^{\mu}(t)$, for high energy atmospheric neutrinos ($E_\nu \sim$ TeV),
\begin{equation}
\begin{split}
& H_0 \sim \frac{\Delta m^2_{\text{atm}}}{2 E_\nu} \sim  10^{-25}~ \text{GeV} ~(\Delta m^2_{\text{atm}} / 10^{-3} ~\text{eV}^2)~ (\text{TeV}/E_{\nu}),\\
&   V_{\text{MSW}}   \simeq 3.8 \times 10^{-14}~\text{eV}~\rho_{\text{matter}}~[\text{g}/\text{cm}^3]\sim 10^{-22}~\text{GeV}~(\rho_{\text{matter}}/10), \\
& a_{\alpha \beta} \simeq \frac{g_{\alpha \beta} |\dot{\phi}_{\odot}|}{M_{Z^{\prime}} } \lesssim 10^{-25}~\text{GeV}, ~~ |a_{\alpha \beta}^{X}| \lesssim 10^{-28}~\text{GeV},
\end{split}
  \end{equation}
due to the presence of the $V_{\text{MSW}}$ and $a_{\alpha \beta}(t),~a_{\alpha \beta}^X(t)$ potentials, propagation of neutrinos can be implemented by dividing the trajectory into $N$-layers of thickness $\Delta L = L/N$. The oscillation probability in the flavor base, with $t_n= t_0+n \Delta L$, given by~\cite{Dev:2020kgz}
\begin{equation}
    P_{\alpha \beta} (t_0, L) = \Big| \bra{\nu_\alpha}  \Big\{\prod_{n=1}^N  \exp [i \tilde{H}_f(t_n) \Delta L]   \Big\} \ket{\nu_\beta} \Big|^2.
\end{equation}
Hence, the observed oscillation probability will be the time average of $P_{\alpha \beta} (t_0, L)$~\cite{Krnjaic:2017zlz} 
\begin{equation}
 \langle P_{\alpha \beta} (L) \rangle  = \frac{1}{\tau_\phi} \int_{0}^{\tau_\phi} dt_0 P_{\alpha \beta} (t_0, L),
\end{equation}
where $\tau_\phi = 2 \pi / m_\phi$ is the period of oscillation of $\phi(t)$.

The corresponding $CPT$-odd LIV isotropic coefficients $(a_{\alpha \beta} \lesssim 10^{-25}~\text{GeV})$~\footnote{Directional-dependent effects $a_{\alpha \beta}^{X}\lesssim 10^{-28}$ GeV (which are four orders of magnitude stronger than current limits~\cite{Kostelecky:2008ts}), can be searched at IceCube and KM3NeT with 0.1$-$100 EeV ultra-high-energy (UHE) neutrinos~\cite{Klop:2017dim}.} can be probed in the near future, for instance at the IceCube and KM3NeT experiments with neutrino energies around 0.1$-$100 PeV~\cite{Ando:2009ts, Klop:2017dim}. For projected sensitivities and some experimental limits on the LIV coefficients, $a_{\alpha \beta}$ and $a_{\alpha \beta}^{X}$, see Tab.~\ref{tab:1}.

Just as importantly, the expected DUNE sensitivity to an ultralight axion-like field coupling with neutrinos is $\tilde{g}_{\alpha \beta} \simeq 10^{-11}-10^{-12}$ eV$^{-1}$~\cite{Huang:2018cwo}; in our proposed scenario, the effective neutrino$-$scalar field coupling $\tilde{g}_{\alpha \beta}  = g_{\alpha \beta} /M_{Z^{\prime}} \lesssim 10^{-13}$ eV$^{-1}~(g_{\alpha \beta}/\lesssim 0.1)~(10^{12}~\text{eV}/ M_{Z^{\prime}})$ which is close to the projected sensitivity at DUNE.

\subsection{Other phenomenological implications}

In addition to the previous considerations, the interaction term in Eq.~(\ref{neutrinoz}) induces a non-standard four-neutrino interaction
\begin{equation}
    H_{\text{NSI}}^{\nu-\nu} = \Tilde{G}_{\alpha \beta} \big( \bar{\nu}_\alpha \gamma^{\mu} (1-\gamma^5) \nu_\beta \big) \big( \bar{\nu}_\alpha \gamma_{\mu} (1-\gamma_5) \nu_\beta \big),
\end{equation}
where the effective coupling $\Tilde{G}_{\alpha \beta} = g_{\alpha \beta}^2/M_{Z^{\prime}}^2 \lesssim 10^{-26}$ eV$^{-2} \lesssim 10^{-3} G_{F}$, the invisible width of the gauge $Z$ boson constraint $\Tilde{G}_{\alpha \beta} \lesssim (1-10)~G_{F}$~\cite{Bilenky:1999dn}. Moreover, neutrino neutral current (NC) non-standard interactions (NSI) with quarks and leptons can be generated via $Z-Z^{\prime}$ mixing, however, suppressed by the mixing angle $\theta_{ZZ^{\prime}} \lesssim 0.1 g_{\alpha \beta}$~\cite{Abdallah:2021npg}
\begin{equation}
    H_{\text{NSI}} \simeq 2 G_F \epsilon_{\alpha \beta}^{f C} \big( \bar{\nu}_\alpha \gamma^{\mu} (1-\gamma^5) \nu_\beta \big) \big( \bar{f}_\alpha \gamma_{\mu} P_C f_\beta \big),
\end{equation}
here $f$ is the corresponding quark or lepton, $ 2 P_C = 1 \pm \gamma_{5},~C = R,L$ is the chiral projector, and $\epsilon_{\alpha \beta} = \sum_{f} \epsilon_{\alpha \beta}^{f C}  \sim 0.1 g_{\alpha \beta}^2 M_Z^2/M_{Z^{\prime}}^2 \lesssim 10^{-5}$. For some experimental limits on the non-standard parameters $\epsilon_{\alpha \beta}$, see Tab.~\ref{tab:1}, bounds from charged current (CC) NSI processes do not apply in this scenario.

On the other hand, the phenomenological Lagrangian in Eq.~(\ref{uldmn}) induces $CPT$ violation in electrons at one loop~\cite{Gu:2005eq} via the weak interaction ($W_\mu-$ neutrino loop)
\begin{equation}
\label{uldmloop}
    \mathcal{L}_{1-\text{loop}} = \tilde{a}_e^{\mu} \Bar{e} \gamma_{\mu}(1- \gamma_5) e, ~~ \tilde{a}_e^{\mu} \simeq g_{e e} \frac{\partial^{\mu} \phi}{M_{Z^{\prime}}} \frac{\alpha M_{Z^{\prime}}^2 }{8 \pi \sin^2 \theta_{W} M_W^2}, 
\end{equation}
here $\alpha$ is the fine-structure constant and $\theta_W$ is the weak mixing angle. Experimental bounds set $\tilde{a}_e^{0} < 5 \times 10^{-25}$ GeV~\cite{Gu:2005eq} and $\tilde{a}_e^{X} <  10^{-25}$ GeV~\cite{Kostelecky:2008ts}, for the isotropic and spatial coefficients, respectively. Therefore, at our reference values, we obtain
\begin{equation}
 \tilde{a}_e^{0} \lesssim 10^{-26}~ \text{GeV} ~ (g_{ee}\lesssim 0.1)~(M_{Z^{\prime}} / \text{TeV}), ~~ \tilde{a}_e^{X} \lesssim 10^{-3} \tilde{a}_e^{0}, 
\end{equation} 
which are within the experimental constraints, as long as $M_{Z^{\prime}} \lesssim$ TeV. 
Furthermore, at earlier times, the corresponding interaction in Eq.~(\ref{uldmloop}) could induce an electron-scalar field back-reaction potential $V(n_e, \dot{\phi})\propto n_e \dot{\phi}$, which is expected to be negligible after electron decoupling.

\begin{table}[H]
\caption{\label{tab:1} Summary of current limits and projected sensitivities (shown in parenthesis) from beyond the standard model: LIV, NC NSI, and neutrino-scalar decay processes relevant to this study.}
\centering
\begin{tabular}{ c   c }
\hline \hline
~BSM process~ & Limit (Sensitivity) \\
\hline 
~LIV neutrino sector~~&~~~$|a_{e \mu}| < 1.8 \times 10^{-23}$ GeV,~Super-Kamiokande~\cite{Super-Kamiokande:2014exs}~ \\ 
~LIV neutrino sector~~&~~~$|a_{e \tau}| < 2.8 \times 10^{-23}$ GeV,~Super-Kamiokande~\cite{Super-Kamiokande:2014exs}~ \\
~LIV neutrino sector~~&$|a_{\mu \tau}| < 2.9 \times 10^{-24}$ GeV,~IceCube~\cite{IceCube:2017qyp} \\
~LIV neutrino sector~~&$a_{\tau \tau} < 2 \times 10^{-26}$ GeV,~IceCube~\cite{Arguelles:2024cjj} \\
~LIV neutrino sector~~&($|a_{\alpha \beta}| \sim 10^{-23}$ GeV),~DUNE~\cite{Barenboim:2018ctx, Agarwalla:2023wft}  \\
~LIV neutrino sector~~&($a_{\alpha \beta} \lesssim 10^{-30}$ GeV),~UHE neutrinos~\cite{Ando:2009ts, Klop:2017dim}\\
~LIV neutrino sector~~&($a_{\alpha \beta}^{X}\lesssim 10^{-30}$ GeV),~UHE neutrinos~\cite{Ando:2009ts, Klop:2017dim}\\
~LIV neutrino sector~~&$a_{\alpha \beta} \lesssim 10^{-25}$ GeV, this work\\
~LIV neutrino sector~~&$|a_{\alpha \beta}^{X}|\lesssim 10^{-28}$ GeV, this work\\
~LIV neutrino sector~~&~($\tilde{g}_{\alpha \beta} \simeq 10^{-11}-10^{-12}$ eV$^{-1}$),~DUNE~\cite{Huang:2018cwo} \\
~LIV neutrino sector~~&~$\tilde{g}_{\tau \tau} < 3 \times 10^{-13}$ eV$^{-1}$,~IceCube~\cite{Arguelles:2024cjj} \\
~LIV neutrino sector~~&~$\tilde{g}_{\alpha \beta} \lesssim 10^{-13}$ eV$^{-1}$,~this work \\
~LIV electron sector~&$\tilde{a}_e^{0} < 5 \times 10^{-25}$ GeV~\cite{Gu:2005eq}\\
~LIV electron sector~&$\tilde{a}_e^{X} <  10^{-25}$ GeV~\cite{Kostelecky:2008ts}\\
~LIV electron sector~&$\tilde{a}_e^{0} \lesssim 10^{-26}$ GeV, this work\\
~LIV electron sector~&$\tilde{a}_e^{X} \lesssim 10^{-29}$ GeV, this work\\
NC NSI~&~$|\epsilon_{\mu \tau}^{qV}| < 5.4 \times 10^{-3}$~\cite{Farzan:2017xzy}~ \\
NC NSI~&~~$-4.1\times 10^{-3} \lesssim \epsilon^{\oplus}_{\mu \tau} \lesssim 3.1 \times 10^{-3}$, IceCube~\cite{IceCube:2022ubv} \\
NC NSI~& $\epsilon_{\alpha \beta}\lesssim 10^{-5}$, this work \\
$\nu_{\alpha} \leftrightarrow \nu_{\beta} + \phi$~&~$\tilde{g}_{\alpha \beta} \lesssim 10^{-10}$ eV$^{-1}$~\cite{Huang:2018cwo}\\
 \hline \hline
\end{tabular}
\end{table} 
Regarding the scalar field $\phi$, radiative corrections could destabilize its mass. For instance, contributions to the scalar field mass $m_\phi \sim 10^{-15}$ eV can be generated from the effective LIV neutrino interaction in Eq.~(\ref{uldmn}) 
\begin{equation}
  \delta m_\phi \sim \frac{g_{\alpha \beta} m_\phi}{ \pi M_{Z^{\prime}}} \Lambda \lesssim m_\phi ~(g_{\alpha \beta}/\lesssim \pi)~(\Lambda/M_{Z^{\prime}}).
\end{equation}
Besides, the corresponding Lagrangian in Eq.~(\ref{uldmn}) may produce scattering between the neutrinos and the scalar particles. However, these interactions are expected to be negligible~\cite{Klop:2017dim}.

Moreover, the neutrino current in Eq.~(\ref{neutrinoz}) induces radiative corrections to the neutrino mass 
\begin{equation}
    \delta m_\nu \sim \frac{g_{\alpha \beta}^2 m_\nu}{\pi^2 M_{{Z}^{\prime }}^2} \Lambda^2 \lesssim 10^{-3} m_\nu ~(g_{\alpha \beta}/ \lesssim 0.1)~ (\Lambda/M_{{Z}^{\prime}}). 
\end{equation}

\section{Conclusions}
\label{conclusion}
In the cosmological context, scalar fields have been used to describe the behavior of dark matter and dark energy. The interaction of these fields with matter could be an important way to determine its existence by searching for its possible observational and experimental consequences.

In this paper, we have studied the interactions between the active neutrinos with a time-varying scalar field, which might lead to possible signals of Lorentz invariance and $CPT$ violation in future and present neutrino oscillation experiments such as IceCube and the KM3NeT proposal. Furthermore, we discussed how those interactions induce a back-reaction potential as well as energy exchange among the neutrinos and the scalar field, modifying its cosmological evolution. The scalar field can act as either cosmological dark energy or dark matter, depending on the form of the neutrino-scalar field energy exchange term.

In addition, we outlined some phenomenological consequences and constraints in the scenario where the scalar field $\phi$ can be considered a dark matter candidate.

\section*{Acknowledgments}
We would like to acknowledge O.~G.~Miranda for insightful discussions and suggestions. This work was partially supported by SNII-M\'exico and CONAHCyT research Grant No.~A1-S-23238. Additionally, the work of R.C. was partially supported by COFAA-IPN, Estímulos al Desempeño de los Investigadores (EDI)-IPN and SIP-IPN Grants No.~20221329, No.~20230732 and No.~20241624. We acknowledge the anonymous referee for the illuminating comments and suggestions.

\appendix

\section{Boson current in a non-linear realization}
\label{reali}
The interaction term from Eq.~(\ref{uldmz}), could arise from a non-linear CCWZ realization~\cite{Coleman:1969sm, Callan:1969sn,Bando:1987br,Kitazawa:2022gzk}, in analogy with that of pions and kaons. For instance, considering a non-linear Nambu-Goldstone field
\begin{equation}
    \Phi = e^{-i \phi/ f_\phi},
\end{equation}
being $\phi$ a real scalar field and $f_\phi$ some energy scale, with an effective Lagrangian
\begin{equation}
\mathcal{L}_{\Phi,Z^{\prime}}^{\text{int}} \supset \frac{f_\phi^2}{2}\mathcal{D}_\mu \Phi \mathcal{D}^\mu \Phi -V(\Phi) +g_\phi f_\phi^2  \mathcal{D}^\mu \Phi Z_\mu^{\prime},
\end{equation}
where the corresponding derivative $f_\phi \mathcal{D}_\mu \Phi$ is given by~\cite{Bando:1987br}
\begin{equation}
   f_\phi \mathcal{D}^{\mu} \Phi = if_{\phi} \big(e^{i \phi/f_{\phi}}\partial^{\mu} e^{-i \phi/f_{\phi}} \big)=\partial^{\mu} \phi +\frac{1}{3!} \big(\frac{i}{f_\phi}\big)^2  \big[\phi, [\phi, \partial^{\mu} \phi]\big] + \cdots,
\end{equation}
assuming a potential $V(\Phi)$ quadratic in $\Phi$ 
\begin{equation}
    V(\Phi) = \frac{\mu^4_\phi}{2} \Big[1-\frac{1}{2}\big(\Phi\Phi +h.c. \big) \Big],
\end{equation}
the effective interacting Lagrangian, including the scalar field $\phi$ and vector boson $Z_\mu^{\prime}$ is
\begin{equation}
\label{eqa5}
\mathcal{L}_{\phi,Z^{\prime}}^{\text{int}} \supset \frac{1}{2}\partial_\mu \phi \partial^\mu \phi - \frac{\mu^4_\phi}{2} \Big[1-\cos\big(2 \frac{\phi}{f_\phi} \big)\Big] +g_\phi f_\phi  \partial^\mu \phi Z_\mu^{\prime},
\end{equation}
with periodicity for the scalar field $\phi \simeq \phi+n \pi f_\phi$. The minimum of the scalar field potential $V(\phi)$ is given by
\begin{equation}
    \frac{d V(\phi)}{d \phi} =0 \rightarrow \phi_{\text{min}} = n \pi f_\phi, ~n = 0, 1, 2, \cdots~.
\end{equation}
For $\phi \ll f_\phi$, we can expand the scalar field potential as
\begin{equation}
  V (\phi)  = \frac{\mu^4_\phi}{2} \Big[1-\big( 1 -2 \frac{\phi^2}{f_\phi^2} + \frac{2}{3} \frac{\phi^4}{f_\phi^4} - \cdots \big) \Big] = m_\phi^2 \phi^2 -\frac{1}{3} \lambda_\phi \phi^4 + \cdots,
\end{equation}
 and the dominant term is the mass term $m_\phi =\mu_{\phi}^2/f_\phi$. Hence, for a scalar field with mass $m_\phi\sim 10^{-15}$ eV and $f_\phi \sim 10^{25}$ eV, the quartic coupling is $\lambda_\phi \sim m_\phi^2 f_\phi^{-2} \sim 10^{-80}$. Therefore, a free scalar field was roughly $\phi \sim f_\phi$ at early epochs until $H \sim m_\phi$.
 
In some models, $\mu_{\phi}^4 \sim \tilde{M}_\text{Pl}^2 \Lambda^2 e^{-S}$ ($\tilde{M}_\text{Pl} \sim 10^{18}$ GeV)~\cite{Marsh:2015xka,Hui:2016ltb}, typically $\Lambda / \text{GeV} \sim (10^4, 10^{11}, 10^{18})$ and $S \lesssim 200-300$~\cite{Hui:2016ltb}, thus $S\simeq 170~(\Lambda/ 10^{11} \text{GeV})$. 
 
Regarding the interaction term from Eq.~(\ref{uldmz}), by considering a $Z^{\prime}_{\mu}$ mass $M_{Z^{\prime}} \sim 10^{12}~\text{eV}$ with coupling $ g_\phi \sim \mathcal{O}( M_{Z^{\prime}} / f_\phi) \sim 10^{-13}~( 10^{25}~\text{eV} /f_\phi)$ in Eq.~\ref{eqa5}, leads to the effective interaction term $\mathcal{L}_{\phi, Z^{\prime}} \supset M_{Z^{\prime}}\partial^{\mu} \phi Z_\mu^{\prime}$.

\section{Scalar field evolution}
\label{scalarevo}

The equation of motion for the scalar field, including the neutrino$-$scalar field energy exchange $Q\propto \ddot{\phi}$~\cite{Simpson:2016gph} is
\begin{equation}
\label{eqmotintapx}
    \ddot{\phi}\Big(1+ \frac{g_{\alpha \beta} n_\nu }{M_{Z^{\prime}} \dot{\phi}}  \Big) + 3H \dot{\phi} + \frac{d V(\phi)}{d \phi} =0.
\end{equation}
Besides, in the radiation epoch, the Hubble parameter $H$ is related to the temperature $T$ as
\begin{equation}
H=\frac{1}{2 t} \simeq 1.66 \sqrt{g_{*}} \frac{T^{2}}{M_{\text{Pl}}},
\end{equation}
being $g_{*}$ the effective number of relativistic degrees of freedom and
\begin{equation}
T=\frac{K}{a} . \,\,
\end{equation}
Some aspects of the evolution of the scalar field can be analyzed more conveniently in terms of the temperature, then the following relations are useful: 
\begin{equation}
\begin{split}
&\frac{d T}{d t}=-\frac{K}{a^{2}} \dot{a}=-\frac{K}{a} H=-T H, \\
&\dot{\phi}  =- H T \frac{d \phi}{d T}, \, \,
\end{split}
\end{equation}
and
\begin{equation}
\ddot{\phi}  =3 H^{2} T \frac{d \phi}{d T}+H^{2} T^{2} \frac{d^{2} \phi}{d T^{2}}.
\end{equation}
Now, the interacting equation of motion~\ref{eqmotintapx} with potential $2 V(\phi) = m_\phi^2 \phi^2$ is
\begin{equation}
\label{eqb7}
\ddot{\phi}\left(1+\frac{g_{\alpha \beta} n_\nu }{M_{Z^{\prime}} \dot{\phi}} \right)+3 H \dot{\phi}+m_{\phi}^{2} \phi=0,
\end{equation}
which can be written in terms of the temperature with the help of the previous relations (in Eq.~\ref{eqb7}, instead of the neutrino number density $n_\nu \sim 0.1~T^3$, a term such as $\rho_\nu m_\nu^{-1}$, could be of phenomenological interest as well) in the following form
\begin{equation}
\begin{split}
&\left(1-\frac{0.1 g_{\alpha \beta} M_{\text{Pl}} }{(1.66) \sqrt{g_{*}} M_{Z^{\prime}}  \frac{d \phi}{d T}}\right) \frac{d^{2} \phi}{d T^{2}}+\frac{m_{\phi}^{2} M_{\text{Pl}}^{2} \phi}{(1.66)^{2} g_{*} T^{6}}= 
\frac{(0.3) g_{\alpha \beta} M_{\text{Pl}}}{M_{Z^{\prime}} (1.66) \sqrt{g_{*}}T}. \end{split}
\end{equation}
This expression can be written in the form
\begin{equation}
\begin{split}
\frac{d^{2} \phi}{d T^{2}}-\frac{0.1 g_{\alpha \beta} M_{\text{Pl}} }{(1.66) \sqrt{g_{*}} M_{Z^{\prime}}} \frac{d}{d T}\left(\ln \left(\frac{d \phi}{d T}\right)\right)+\frac{m_{\phi}^{2} M_{\text{Pl}}^{2} \phi}{(1.66)^{2} g_{*} T^{6}}= 
\frac{(0.3) g_{\alpha \beta} M_{\text{Pl}}}{M_{Z^{\prime}} (1.66) \sqrt{g_{*}} T}.
\end{split}
\end{equation}
An alternative expression can be written as
\begin{equation}
\begin{split}
 \left(\frac{d \phi}{d T}-\frac{0.1 g_{\alpha \beta} M_{\text{Pl}} }{(1.66) \sqrt{g_{*}} M_{Z^{\prime}} } \right) \frac{d^{2} \phi}{d T^{2}}= \left(\frac{(0.3) g_{\alpha \beta} M_{\text{Pl}}}{M_{Z^{\prime}} (1.66) \sqrt{g_{*}} T}-\frac{m_{\phi}^{2} M_{\text{Pl}}^{2} \phi}{(1.66)^{2} g_{*} T^{6}}\right) \frac{d \phi}{d T}.
\end{split}
\end{equation}
The former expression is suitable to extract information about the dynamics of the scalar field. When $\frac{d \phi}{d T}=0$, then  $\frac{d^{2} \phi}{d T^{2}}=0$. Moreover, if we derive the former equation, we get
\begin{equation}
\begin{split}
& \frac{d}{d T}\left(\frac{d \phi}{d T}-\frac{0.1 g_{\alpha \beta} M_{\text{Pl}} }{(1.66) \sqrt{g_{*}} M_{Z^{\prime}} }\right) \frac{d^{2} \phi}{d T^{2}}+\left(\frac{d \phi}{d T}-\frac{0.1 g_{\alpha \beta} M_{\text{Pl}} }{(1.66) \sqrt{g_{*} }M_{Z^{\prime}} }\right) \frac{d^{3} \phi}{d T^{3}} \\
& = \frac{d}{d T}\left(\frac{(0.3) g_{\alpha \beta} M_{\text{Pl}}}{M_{Z^{\prime}} (1.66) \sqrt{g_{*}}T}-\frac{m^{2}_\phi M_{\text{Pl}}^{2} \phi}{(1.66)^{2} g_{*} T^{6}}\right) \frac{d \phi}{d T}~ +\\
&~~ \left(\frac{(0.3) g_{\alpha \beta} M_{\text{Pl}}}{M_{Z^{\prime}}(1.66) \sqrt{g_{*}}T}-\frac{m^{2}_\phi M_{\text{Pl}}^{2} \phi}{(1.66)^{2} g_{*} T^{6}}\right) \frac{d^{2} \phi}{d T^{2}},
\end{split}
\end{equation}
then $ \frac{d^{3} \phi}{d T^{3}}=0$ when $\frac{d \phi}{d T}=0$.
In general $\frac{d^{n} \phi}{d T^{n}}=0$ once the scalar field has $\frac{d \phi}{d T}=0$. The last relations may imply a frozen scalar field, $\phi=$ constant, which is a possible solution to the differential equation as well. It would be interesting to study a possible mechanism to escape from this state employing quantum fluctuations or perturbations of the scalar field.


\end{document}